\PassOptionsToPackage{unicode}{hyperref}
\PassOptionsToPackage{hyphens}{url}
\PassOptionsToPackage{dvipsnames,svgnames,x11names}{xcolor}
\documentclass[
]{article}
\usepackage{amsmath,amssymb}
\usepackage{lmodern}
\usepackage{iftex}
\ifPDFTeX
  \usepackage[T1]{fontenc}
  \usepackage[utf8]{inputenc}
  \usepackage{textcomp} 
\else 
  \usepackage{unicode-math}
  \defaultfontfeatures{Scale=MatchLowercase}
  \defaultfontfeatures[\rmfamily]{Ligatures=TeX,Scale=1}
\fi
\IfFileExists{upquote.sty}{\usepackage{upquote}}{}
\IfFileExists{microtype.sty}{
  \usepackage[]{microtype}
  \UseMicrotypeSet[protrusion]{basicmath} 
}{}
\makeatletter
\@ifundefined{KOMAClassName}{
  \IfFileExists{parskip.sty}{%
    \usepackage{parskip}
  }{
    \setlength{\parindent}{0pt}
    \setlength{\parskip}{6pt plus 2pt minus 1pt}}
}{
  \KOMAoptions{parskip=half}}
\makeatother
\usepackage{xcolor}
\usepackage{color}
\usepackage{fancyvrb}

\DefineVerbatimEnvironment{Highlighting}{Verbatim}{commandchars=\\\{\}}
\newenvironment{Shaded}{}{}

\newcommand{\ImportTok}[1]{\textcolor[rgb]{0.00,0.50,0.00}{\textbf{#1}}}

\newcommand{\NormalTok}[1]{#1}
\newcommand{\OperatorTok}[1]{\textcolor[rgb]{0.40,0.40,0.40}{#1}}

\newcommand{\StringTok}[1]{\textcolor[rgb]{0.25,0.44,0.63}{#1}}

\usepackage{graphicx}
\makeatletter
\def\maxwidth{\ifdim\Gin@nat@width>\linewidth\linewidth\else\Gin@nat@width\fi}
\def\maxheight{\ifdim\Gin@nat@height>\textheight\textheight\else\Gin@nat@height\fi}
\makeatother
\setkeys{Gin}{width=\maxwidth,height=\maxheight,keepaspectratio}
\makeatletter
\def\fps@figure{htbp}
\makeatother
\providecommand{\tightlist}{%
  \setlength{\itemsep}{0pt}\setlength{\parskip}{0pt}}
\NewDocumentCommand\citeproctext{}{}
\NewDocumentCommand\citeproc{mm}{%
  \begingroup\def\citeproctext{#2}\cite{#1}\endgroup}
\makeatletter
 \let\@cite@ofmt\@firstofone
 \def\@biblabel#1{}
 \def\@cite#1#2{{#1\if@tempswa , #2\fi}}
\makeatother
\newlength{\cslhangindent}
\newlength{\csllabelwidth}
\newenvironment{CSLReferences}[2] 
 {\begin{list}{}{%
  \setlength{\itemindent}{0pt}
  \setlength{\leftmargin}{0pt}
  \setlength{\parsep}{0pt}
  \ifodd #1
   \setlength{\leftmargin}{\cslhangindent}
   \setlength{\itemindent}{-1\cslhangindent}
  \fi
  \setlength{\itemsep}{#2\baselineskip}}}
 {\end{list}}
\usepackage{calc}

\ifLuaTeX
\usepackage[bidi=basic]{babel}
\else
\usepackage[bidi=default]{babel}
\fi
\babelprovide[main,import]{american}

\def\languageshorthands#1{}
\ifLuaTeX
  \usepackage{selnolig}  
\fi
\IfFileExists{bookmark.sty}{\usepackage{bookmark}}{\usepackage{hyperref}}
\IfFileExists{xurl.sty}{\usepackage{xurl}}{} 
\hypersetup{
  pdftitle={dfreproject: A Python package for astronomical
reprojection},
  pdfauthor={Carter Lee Rhea, Pieter Van Dokkum, Steven R. Janssens,
Imad Pasha, Roberto Abraham, William P Bowman, Deborah Lokhorst, Seery
Chen},
  pdflang={en-US},
  colorlinks=true,
  linkcolor={Maroon},
  filecolor={Maroon},
  citecolor={Blue},
  urlcolor={Blue},
  pdfcreator={LaTeX via pandoc}}

\title{dfreproject: A Python package for astronomical reprojection}

\definecolor{c53baa1}{RGB}{83,186,161}
\definecolor{c202826}{RGB}{32,40,38}

\usepackage[affil-it]{authblk}
\usepackage{orcidlink}
\author[1,2%
  *%
  ]{Carter Lee Rhea%
    \,\orcidlink{0000-0003-2001-1076}\,%
    }
\author[1,3%
  *%
  ]{Pieter Van Dokkum%
    }
\author[1%
  ]{Steven R. Janssens%
    \,\orcidlink{0000-0003-0327-3322}\,%
    }
\author[1,3%
  ]{Imad Pasha%
    }
\author[1,4,5%
  ]{Roberto Abraham%
    }
\author[1,3%
  ]{William P Bowman%
    \,\orcidlink{0000-0003-4381-5245}\,%
    }
\author[1,6%
  ]{Deborah Lokhorst%
    }
\author[1,4,5%
  ]{Seery Chen%
    }

\affil[1]{Dragonfly Focused Research Organization, 150 Washington
Avenue, Santa Fe, 87501, NM, USA%
  }
\affil[2]{Centre de recherche en astrophysique du Québec (CRAQ)%
  }
\affil[3]{Astronomy Department, Yale University, 219 Prospect St, New
Haven, CT 06511, USA%
  }
\affil[4]{David A. Dunlap Department of Astronomy \& Astrophysics,
University of Toronto, 50 St.~George Street, Toronto, ON M5S3H4, Canada%
  }
\affil[5]{Dunlap Institute for Astronomy \& Astrophysics, University of
Toronto, 50 St.~George Street, Toronto, ON M5S3H4, Canada%
  }
\affil[6]{NRC Herzberg Astronomy \& Astrophysics Research Centre, 5071
West Saanich Road, Victoria, BC V9E 2E7, Canada%
  }
\affil[*]{These authors contributed equally.}
\date{01 April 2025}

\begin{document}
\maketitle

\section{1. Summary}\label{summary}

Deep astronomical images are often constructed by digitially stacking
many individual sub-exposures. Each sub-exposure is expected to show
small differences in the positions of stars and other objects in the
field, due to the movement of the celestial bodies,
changes/imperfections in the opto-mechanical imaging train, and other
factors. To maximize image quality, one must ensure that each
sub-exposure is aligned to a common frame of reference prior to
stacking. This is done by reprojecting each exposure onto a common
target grid defined using a World Coordinate System (WCS) that is
defined by mapping the known angular positions of reference objects to
their observed spatial positions on each image. The transformations
needed to reproject images involve complicated trigonometric expressions
which can be slow to compute, so reprojection can be a major bottleneck
in image processing pipelines.

To make astronomical reprojections faster to implement in pipelines, we
have written \texttt{dfreproject}, a Python package of GPU-optimized
functions for this purpose. The package's functions break down
coordinate transformations using Gnomonic projections to define
pixel-by-pixel shifts from the source to the target plane. The package
also provides tools for interpolating a source image onto a target plane
with a single function call. This module follows the FITS and SIP
formats laid out in the following papers: Greisen \& Calabretta
(\citeproc{ref-greisen_representations_2002}{2002}), Calabretta \&
Greisen (\citeproc{ref-calabretta_representations_2002}{2002}), and
Shupe et al. (\citeproc{ref-shupe_sip_2005}{2005}). Compared to common
alternatives, \texttt{dfreproject}'s routines result in speedups of up
to 20X when run on a GPU and 10X when run on a CPU.

\section{2. Statement of need}\label{statement-of-need}

Several packages already exist for calculating and applying the
reprojection of a source image onto a target plane, such as
\texttt{reproject} (\citeproc{ref-robitaille_reproject_2020}{Robitaille
et al., 2020}) or \texttt{astroalign}
(\citeproc{ref-beroiz_astroalign_2020}{Beroiz et al., 2020}). These
packages excel at easy-to-use, general-purpose astronomical image
reprojection, but they function solely on CPUs and can be computational
bottlenecks in some data reduction pipelines. The \texttt{dfreproject}
package harnesses GPUs (using \texttt{PyTorch}
(\citeproc{ref-paszke_pytorch_2019}{Paszke et al., 2019}) as its
computational backbone) to improve computational efficiency. The package
has minimal reliance on pre-existing WCS packages such as those found in
\texttt{astropy} or \texttt{WCSLIB}
(\citeproc{ref-astropy_collaboration_astropy_2013}{Astropy Collaboration
et al., 2013}, \citeproc{ref-astropy_collaboration_astropy_2018}{2018},
\citeproc{ref-astropy_collaboration_astropy_2022}{2022}), with such
packages being used only for convenience in non-computationally
expensive steps (such as using \texttt{astropy.wcs} to read the header
information from the input fits files).

\texttt{dfreproject}'s primary purpose is to reproject observations
taken by a new version of the Dragonfly Telephoto Array that is
currently under construction in Chile. The volume of data obtained with
this telescope will be large, with \textgreater1000 exposures obtained
simultaneously, and it is paramount that the processing pipeline
incorporates fast and accurate reprojection methods. As shown below, by
leveraging \texttt{PyTorch} for vectorization and parallelization via
the GPU, \texttt{dfreproject} achieves a considerable speedup (up to
nearly 20X) over standard methods.

\texttt{dfreproject} can be used as a direct replacement for
\texttt{reproject.reproject\_interp} by simply importing
\texttt{dfreproject} instead of \texttt{reproject} such as:

\begin{Shaded}
\begin{Highlighting}[]
\ImportTok{from}\NormalTok{ dfreproject }\ImportTok{import}\NormalTok{ calculate\_reprojection}
\NormalTok{reprojected }\OperatorTok{=}\NormalTok{ calculate\_reprojection(}
\NormalTok{    source\_hdus}\OperatorTok{=}\NormalTok{source\_hdu,}
\NormalTok{    target\_wcs}\OperatorTok{=}\NormalTok{target\_wcs,}
\NormalTok{    shape\_out}\OperatorTok{=}\NormalTok{target\_hdu.data.shape,}
\NormalTok{    order}\OperatorTok{=}\StringTok{\textquotesingle{}bilinear\textquotesingle{}}
\NormalTok{)}
\end{Highlighting}
\end{Shaded}

The \texttt{target\_wcs} argument can be passed as a header similar to
\texttt{reproject.} Additionally, if \texttt{shape\_out} is not
provided, the shape will be the same as the input.

\section{3. Methods}\label{methods}

\subsection{3.1 Overview}\label{overview}

We must perform three intermediate calculations to reproject an image
onto a new coordinate plane. To do this, we use the target and source
WCS.

Before defining the steps, there are a few terms to define:

\begin{itemize}
\tightlist
\item
  SIP: Standard Imaging Polynomial. This convention allows us to
  represent non-linear geometric distortions as a simple polynomial. The
  order and coefficients of this polynomial are stored in the header.
  The SIP is broken down into four individual polynomials, SIP\_A,
  SIP\_B, SIP\_INV\_A, and SIP\_INV\_B where SIP\_A defines the
  polynomial applied to the x-coordinates, SIP\_B defines the polynomial
  applied to the y-coordinates, and SIP\_INV\_A and SIP\_INV\_B define
  the inverse operations. For an in-depth discussion on SIP, please see
  Shupe et al. (\citeproc{ref-shupe_sip_2005}{2005}).
\end{itemize}

CD Matrix: Coordinate Description Matrix. This is a 2X2 matrix that
encodes the image's rotation, skew, and scaling. The values are
conveniently stored in the header. The CD matrix may also be constructed
from the PC Projection Coordinate matrix multiplied by the CDELT values.

The steps are as follows:

\begin{enumerate}
\def\labelenumi{\arabic{enumi}.}
\item
  For each pixel, calculate the corresponding celestial coordinate using
  the target WCS

  \begin{enumerate}
  \def\labelenumii{\arabic{enumii}.}
  \item
    Apply shift
  \item
    Apply SIP distortion
  \item
    Apply CD matrix
  \item
    Apply transformation to celestial coordinates using the Gnomonic
    projection
  \end{enumerate}
\item
  Calculate the position in the source grid for each celestial
  coordinate. This provides the offset for the next step.

  \begin{enumerate}
  \def\labelenumii{\arabic{enumii}.}
  \item
    Apply the inverse transformation using the Gnomonic projection
  \item
    Apply inverse CD matrix
  \item
    Apply inverse SIP distortion
  \item
    Apply shift
  \end{enumerate}
\item
  Interpolate the source image onto the newly calculated grid
\end{enumerate}

In the final interpolation step, we include local flux conservation by
simultaneously projecting an identity tensor called the footprint. The
final reprojected frame is normalized by this footprint.

\subsection{3.2 Coordinate
Transformation}\label{coordinate-transformation}

In this section, we describe the coordinate transformation using the
Gnomonic projection. Please note that we include an additional shift of
1 pixel to handle Python being 0-based. We will be using the following
definitions for values:

\(x,y\) - pixel values

\(\mathrm{crpix}_1, \mathrm{crpix}_2\) - center pixels as defined in WCS

\(\mathrm{dec}_0, \mathrm{ra}_0\) - Central Declination and Right
Ascension as defined in the WCS.

All trigonometric functions require the values to be in radians.

\subsubsection{3.2.1 To celestial
coordinates}\label{to-celestial-coordinates}

For these calculations, we use the WCS information for the target plane.

\[u = x - (\mathrm{crpix}_1 - 1) \] \[v = y - (\mathrm{crpix}_2 -1) \]

\[u = u - \mathrm{SIP\_A}(u, v)\] \[v = v - \mathrm{SIP\_B}(u,v)\]

\[
\begin{bmatrix}
u' \\
v'
\end{bmatrix}
 = \mathrm{CD}
\begin{bmatrix}
u \\
v
\end{bmatrix}
\] \[r = \sqrt{u'^2 + v'^2}\] \[r_0 = \frac{180}{\pi}\]
\[\phi = \mathrm{atan}^2(-u', v')\]
\[\theta = \mathrm{atan}^2(r_0, r) \]
\[\mathrm{dec} = \sin^{-1}\Big( \sin(\theta)\sin(dec_0) + \cos(\theta)\cos(dec_0)\cos(\phi) \Big) \]
\[\mathrm{ra} = \mathrm{ra}_0 + \mathrm{atan}^2\Big( -\cos(\theta)\sin(\phi), \sin(\theta)\cos(dec_0)-\cos(\theta)\sin(dec_0)\cos(\phi) \Big) \]

\subsubsection{3.2.2 To source coordinates}\label{to-source-coordinates}

For these calculations, we use the WCS information for the source plane.

\[\Delta \mathrm{ra} = \mathrm{ra} - \mathrm{ra}_0\]
\[\phi = \mathrm{atan}^2\Big(-\cos(\mathrm{dec})\sin(\Delta \mathrm{ra}), \sin(\mathrm{dec})\cos(\mathrm{dec}_0)-\cos(\mathrm{dec})\sin(\mathrm{dec}_0)\cos(\Delta \mathrm{ra}) \Big) \]
\[\theta = \sin^{-1}\Big( \sin(\mathrm{dec})\sin(\mathrm{dec}_0) + \cos(\mathrm{dec})\cos(\mathrm{dec}_0)\cos(\Delta \mathrm{ra}) \Big) \]
\[r = r_0\frac{\cos(\theta)}{\sin(\theta)}\] \[u' = -r\sin(\phi) \]
\[v' = r\cos(\phi)\]

\[
\begin{bmatrix}
u \\
v
\end{bmatrix}
 = \mathrm{CD\_INV}
\begin{bmatrix}
u' \\
v'
\end{bmatrix}
\]

\[u = u - \mathrm{SIP\_INV\_A}(u, v)\]
\[v = v - \mathrm{SIP\_INV\_B}(u, v)\]

\[x = u + (\mathrm{crpix}_1 - 1)\] \[y = v + (\mathrm{crpix}_2 - 1)\]

\section{4. Results}\label{results}

\subsection{4.1 Demo}\label{demo}

\begin{figure}
\centering
\includegraphics{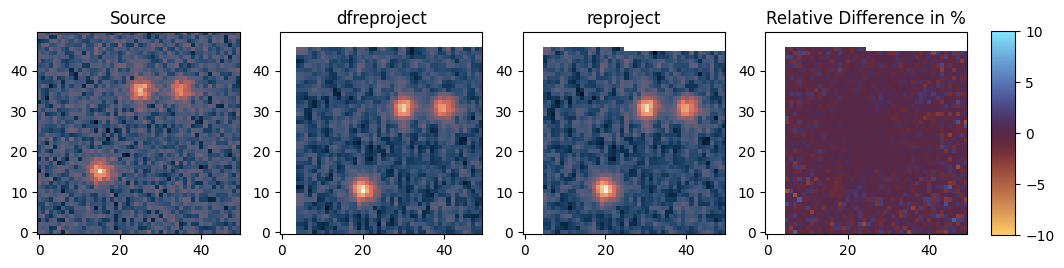}
\caption{\label{fig:demo}}
\end{figure}

We created two small (50x50) FITS files for this demonstration with a
pixel offset of 0.5 pixels and a 0.005 degree offset in the heeader. In
\autoref{fig:demo} from left to right, we show the original image, the
\texttt{dfreproject} solution, the \texttt{reproject} solution, and the
relative error between the two. We define the relative error as
\(100 * \frac{\mathrm{dfreproject\_solution} - \mathrm{reproject\_solution}}{\mathrm{reproject\_solution}}\).
For both solutions, we use a bilinear interpolation scheme. In the noisy
regions of the image, the differences in the reprojections is pure
noise. There are slight differences in the solutions at the locations of
the Gaussians which is attributable to small differences in the
normalization.

\subsection{4.2 Speed Comparison}\label{speed-comparison}

\begin{figure}
\centering
\includegraphics{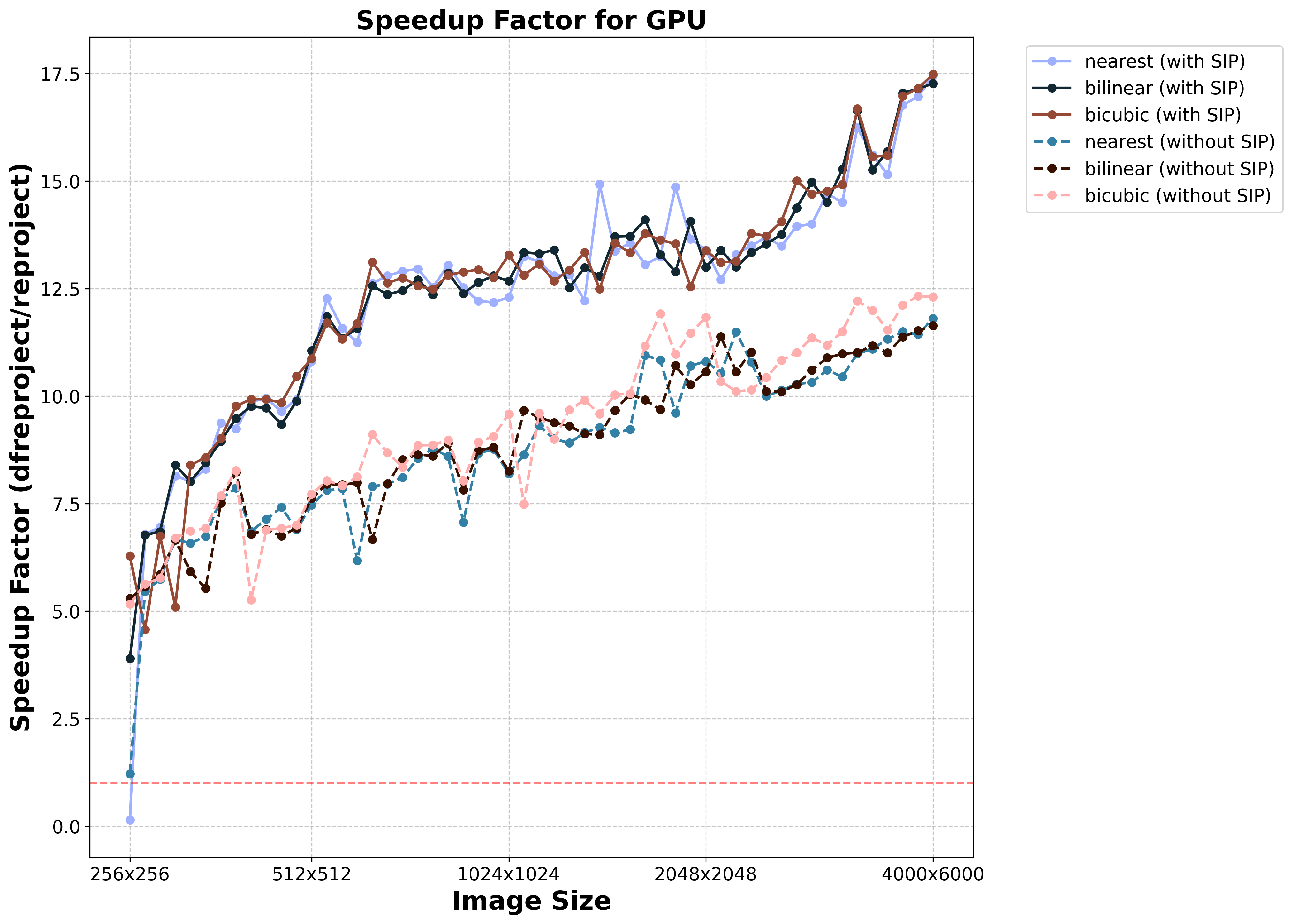}
\caption{\label{fig:gpu-comparison}}
\end{figure}

To compare the execution times, we created a benchmarking script (which
can be found in the demos/benchmarking directory under
\texttt{benchmark-script.py}; the figures are constructed with
\texttt{benchmark-plotter.py}). This test is run between
\texttt{dfreproject} and \texttt{reproject}. We benchmark the three
interpolation schemes with and without SIP distortion for images sized
256x256 to 4000x6000 (this approximately matches the size of Dragonfly
images). \autoref{fig:gpu-comparison} shows the results of this
benchmarking when \texttt{dfreproject} is run using a GPU (NVIDIA
GeForce RTX 4060).

As evidenced by this figure, \texttt{dfreproject} has a significant
speed advantage over \texttt{reproject} for larger images regardless of
the type of interpolation scheme. The speedup is most pronounced in the
case where SIP distortions are included.

\begin{figure}
\centering
\includegraphics{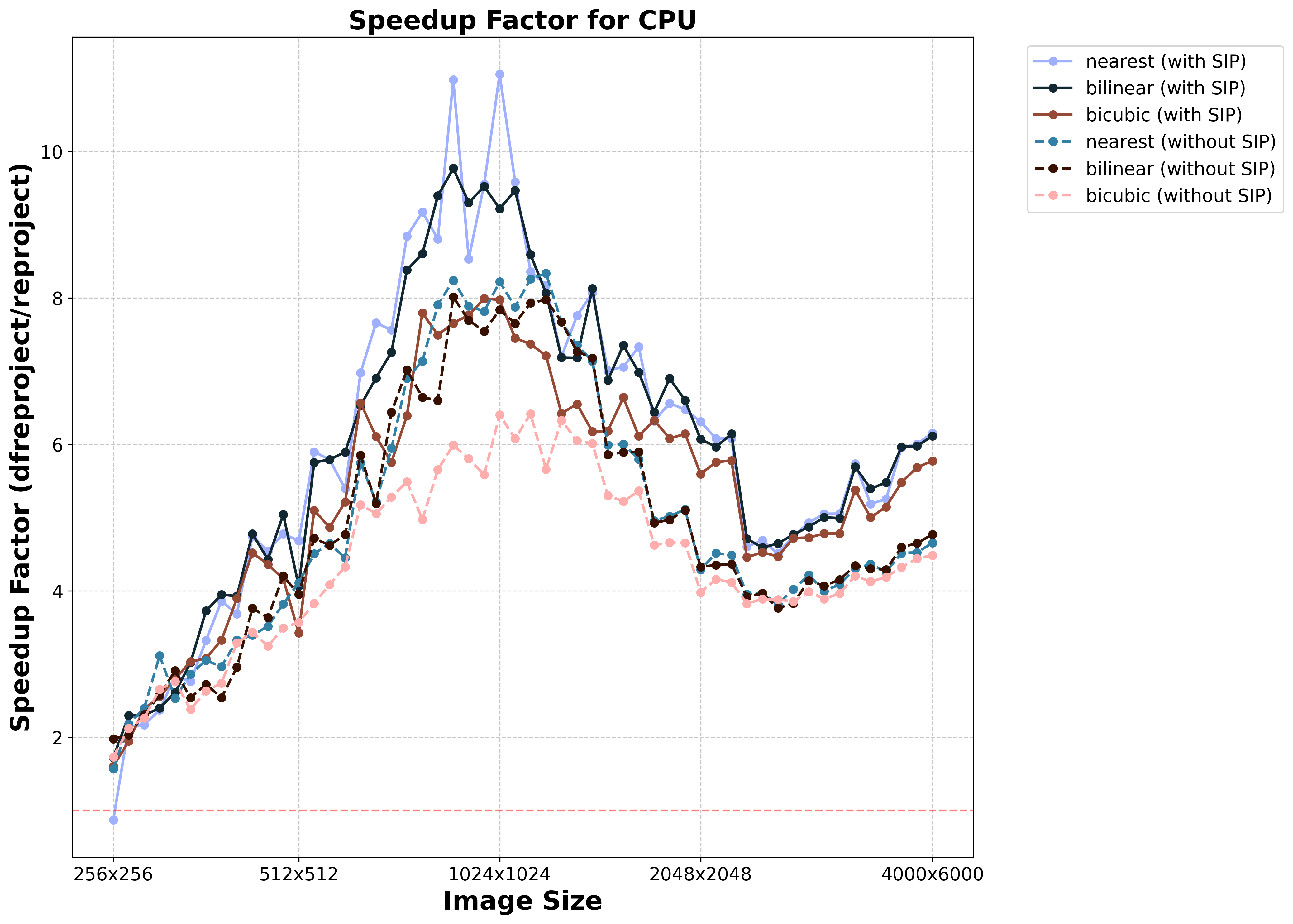}
\caption{\label{fig:cpu-comparison}}
\end{figure}

In \autoref{fig:cpu-comparison}, we display the same results except we
used a CPU (Intel® Core™ i9-14900HX).

Although the speedup on the CPU is not as impressive as on the GPU, it
is still considerable.

All code can be found in the \texttt{demo} directory.

\section{Acknowledgements}\label{acknowledgements}

We acknowledge the Dragonfly FRO and particularly thank Lisa Sloan for
her project management skills.

We use the cmcrameri scientific color maps in our demos
(\citeproc{ref-crameri_scientific_2023}{Crameri, 2023}).

\section*{References}\label{references}
\addcontentsline{toc}{section}{References}

\phantomsection\label{refs}

\end{document}